\documentstyle{lamuphys}
\makeatletter
\let\chapter\hid@chapter
\makeatother
\begin{document}
\pagenumbering{arabic}
\titlerunning{Dust in Hot Environments}
\title{Dust in Hot Environments: Giant Dusty Galactic Halos}

\author{Andrea Ferrara\inst{1}}

\institute{Osservatorio Astrofisico di Arcetri, Largo Fermi 5, 50125 Firenze,
Italy}

\maketitle

\begin{abstract}
I review some of the evidences for dust in the Local Bubble and in galactic
halos and show that a general mechanism based on radiation pressure is capable of
evacuating dust grains from regions dominated by massive star energy input
and thus originate huge dusty halos. A Monte Carlo/particle model has been
developed to study the dust dynamics above HII chimneys and the results, among
other findings, show that dust can travel several kpc away from the plane of 
the parent galaxy. The cosmological implications of extragalactic dust are
briefly outlined. 
\end{abstract}
\section{Introduction}
The evolution of dust in hot environments is a very important topic in
astrophysics with a variety of implications. There are several excellent reviews
on the subject (Draine \& Salpeter 1979, Seab 1987, McKee 1989, Dwek \& Arendt
1992) - and in 
particular on the issue of grain survival - to which I defer the interested
reader for a complete treatment. Instead, in the spirit of a highlight talk,
I would like to focus on and isolate one particular effect, namely the dynamics
of grains in a hot environment, and show how this might affect the structure 
and evolution of a galaxy and perhaps even our understanding of the distant
universe, as other speakers at this Colloquium have emphasized (see 
Finkbeiner's contribution).

In this sense,  three "hot'' environments - the Local Bubble (LB),
Galactic Halos
and the Intergalactic Medium (IGM) - central to the subject of this
Colloquium are
probably different aspects of the same phenomenon: the feedback of star
formation
and supernovae on the interstellar medium of a galaxy,  whose
manifestation is the
disk/halo interaction. The central point of this paper is that the
disk/halo interaction
- and in particular its dusty form - is indeed crucial to understand
several galactic
properties and also the distant universe. The relatively well studied
local environment
might then represent the Rosetta stone to interpret  the interstellar
processes occurring
during the epoch of galaxy formation.

\section{Dust in the Local Bubble}

Contrary to what one might expect, not much is known yet about dust in
the LB. The most firm results  come from a pioneering paper by Tinbergen
(1982).
This author observed that stars closer than $\simeq 35$~pc from the sun in 
the
direction of the Galactic Center show appreciable polarization (polarization
degree of the order of  0.02 \%). From the same data  Frisch (1995) found no
correlation between the polarization degree and the the distance to the sample
stars. This could indicate that  the dust is closer than the nearest star,
and therefore
located in the neutral cloud embedding the sun, commonly named the Local
Fluff. One might also be led to speculate that the amount of dust outside 
this
very small region of a few parsecs size is very limited. Another
interesting feature
is the absence of polarization in the anticenter hemispere, which could be
explained by a lack of grain alignement and/or a different configuration of 
the
magnetic field. It has to be pointed out, thought, that Leroy (1993) did a
very similar
study to the one of Tinbergen, failing to confirm the above discussed
polarization;
however, this could be accounted for by the lower sensitivity of Leroy's data.
New promising perspectives for the investigation of dust in the local
environment
are open by ongoing infrared emission studies (Reach, this conference) and
in-situ
experiments (Landgraf, this conference).

The absence of polarization outside the very local region might have different
interpretations. If one is inclined to attribute this effect to a real
deficiency of
grains rather than to other reasons, such as a poor alignment,  then the issue
of grain destruction or evacuation should be addressed.
Grain in a hot gas are essentially destroyed via thermal sputtering, i.e. 
collisions
with ions or electrons with Maxwellian velocity distribution. The hot gas
in the LB
is now thought to have temperature $\sim 10^6$~K and number density $n\sim
5\times 10^{-3}$~cm$^{-3}$. For this figures, the sputtering time calculated
using the recent sputtering yields by Tielens et al. (1994),
for a $0.01~\mu$m
grain is $\sim 10^8$~yr, much longer than the estimated age of the LB.
Dust destruction can have been caused by the supernova shock that has
generated the hot bubble, whose mechanical luminosity has been inferred
to be equal to $\sim 2.8\times 10^{36}$~ergs~s$^{-1}$ by Frisch (1995).
However, in spite of the still poorly understood underlying physics, it seems
unlikely that the efficiency of grain destruction in a shock can be higher
that 10\%
(McKee 1989). Grains are more likely to be destroyed behind radiative shocks, 
by the combined effects of a greatly enhanced gas density and betatron
acceleration
that increases the grain Larmor frequency. There are clear evidences of
supersonic
motions in cloudlets inside the Local Fluff, and collisions among them
necessarily
produce shocks. Ricotti, Ferrara \& Miniati (1997) have shown that supersonic
cloud-cloud collisions are inelastic for a wide range of parameters,
which, stated
in an alternate manner, implies that the corresponding shocks are radiative;
 this
result has now been confirmed by extensive hydro and MHD numerical simulations
performed by Miniati et al. (1997). Thus, the presence of dust in a 5~pc
sphere around
the sun, confirms that the dust destruction has not been very efficient.
This forms a
basis for our hypothesis that dust can have been {\it evacuated} from the
LB rather
than destroyed locally. If the LB is an example, and probably
far cry from being the most spectacular one (see Normandeau et al. 1996),  of the
effects
of  (multi)--supernova explosions resulting in the vertical tunneling of
hot gas
inside HI walls -- in brief a galactic "chymney''-- we do expect that the
global effects
of a collection of such objects might profoundly influence the structure
and properties
of the ISM of galaxies and of their halos. A very brief summary of some
observational
evidences that might hint at the relevance of supernova energy injection
in shaping the
dust distribution in galaxies is given in the next Section.

\section{Observational Facts on Dust Large Scale Distribution}

The tremendous observational improvement occurred in the last few years
has substantially challenged some of our long-standing prejudices 
about distribution of dust in galaxies. Dust has long been
believed to be confined in a very thin disk, with approximately the same
horizontal extension as disk stars. However, this does not appear to be
the case from the first dedicated observations which tackle the problem
exploiting
the range of wavelengths appropriate for the detection of cold dust, i.e. 
FIR
and sub-mm bands. Davies et al. (1997) from an analysis of  140~$\mu$m and
240~$\mu$m DIRBE observations of the Galaxy  demostrated very convincingly
the existence of an extended (scaleheight $\sim 0.5$~kpc), cool ($T\sim
18-22$~K)
dust component. They favour the interpretation that "...this dust may be
supported
high above the plane by radiation pressure''. Dust therefore seems to be
present
in the lower halo of the Milky Way at least at the same distance as the HI gas
in the Lockman layer. Sofue et al. 1994, pointed out the presence of a
variety of
dust structures (arcs, loops, bubbles, streamers) in the halo of NGC~253
extending
into the first 3~kpc of the halo of that galaxies, this limit merely being
set by instrumental sensitivity. This adds with the
already well established evidences of similar phenomena in several edge-on galaxies (where
the identification
is definitely easier); for a review see Dettmar (1992).
Not only the vertical distribution in spirals is more extented than
previously thought,
but similar conclusions are emerging  also for the one in the disk plane. 
Observations of NGC~6949 in the ISO 200~$\mu$m band have shown that a
cold dust component exists which is considerably more extended compared to
that
measured by IRAS in the same galaxy and sampling a warmer component.
Zaritsky
(1994), studying the B and I colors of distant galaxies seen through the halo of
two nearby spirals, concluded that background galaxies at smaller projected
separations are statistically redder than those in the outer regions. This fact
suggests the existence of an extended dust halo with a derived scale length of
$31 \pm 8$~kpc. This  figure is in remarkable agreement with the prediction of
Heisler \& Ostriker (1988), who required a scale length of 33~kpc to account for
the observed quasar number counts.

\section{Grain Dynamics above Bubbles}

There are two main mechanisms able to evacuate dust from the interior of
a supernova driven interstellar bubble and inject it into the halo:  
(i) a convective flow (i.e. a ``galactic fountain'',  Shapiro \& Field [1976] 
or a chimney, Norman \& Ikeuchi [1989])
in which the gas heated by a supernova
blastwave becomes buoyant and raises into the halo carrying along the dust
grains that are coupled with the gas; (ii) a wind driven by the radiation
pressure above the clusters of young OB stars. In order to work efficiently,
the first mechanism requires the occurence of the so-called ``blow-out'' in
which the gas of the shell created by the explosion is effectively reaccelerated
and injected into the halo. However, recent studies have demonstrated that
this phenomenon, at least in the Galaxy, should be much rarer than previously
thought, either because the growth of the shell is inhibited by the presence of
a magnetic field (Tomisaka 1990; MacLow \& Norman 1992) or by effects related to
non-coeval star formation (Shull \& Saken 1995), which essentially result in
a time dependent mechanical luminosity, in turn producing lower expansion
velocities. Here we explore the second possibility, i.e. dust is injected into 
the intergalactic space by radiation pressure. It is interesting to note that
through this mechanism most of the dust is evacuated form the region surrounding
the star cluster {\it prior} to supernova explosions. As the latter take place, 
only some dust will be still close enough to the center of the star cluster to 
be reached and processed by the expanding shock. Thus, at least part of the
grain population will have a higher chance to survive. This effect will occur no
matter how big the association is (and also for an isolated supernova) since
grains will be always pre-exposed to the radiative flux of the stars eventually
turning into supernovae.

To substantiate the above points and to understand the radiation-driven 
dynamics of grains above these so-called ``HII chimneys'' - vertical, density 
bounded ionization structures (Dove \& Shull 1994) allowing Lyc photons to 
escape the disk and penetrate into the halo -  we have performed a set of
numerical simulations
of the above ``dusty chimneys'' based on a mixed Monte Carlo/particle approach.
A complete description of the calculation and of the detailed results will
be presented in Ferrara \& Shull (1997). Here we present the main features of
the model and discuss some results relevant to the present topic. 

We assume that dust grains are immersed in the time-dependent radiation field of
a stellar association containing $N$ OB stars (typically $N=100$). The spectrum
of the radiation field has been adapted from the results of Sutherland \& Shull
(private communication) who calculate the evolution of the composite radiation spectrum from the
most updated stellar models. The evolution of the radiation spectrum for a 
poor association with $N=40$ is shown in Fig. 1. The luminosity increases up to
a maximum and then decreases and fades away as the stars 
evolve into supernovae. 
At $t=42$~Myr time, the luminosity, $L_\nu$,
of the radiation field at 5~eV is roughly 3
orders of magnitude lower than the maximum intensity $L_\nu \sim
10^{24}$~erg~s$^{-1}$~
Hz$^{-1}$~sr$^{-1}$,
reached at $t\sim 17$~Myr. 
The spectrum is characterized by a Lyc break of about
1.5 decades, and by a flat distribution below the Lyman limit. Our simulations
end at $t\sim 50$~Myr, when there is essentially no power in the radiations to drive
the grains.

\begin{figure}
\vspace{7.0cm}
\caption{Time evolution of the radiation spectrum above an OB
association containing $N=40$ OB stars as a function of photon energy. 
From top to bottom the curves refer to the evolutionary times $17, 29, 1, 42$~Myr,
respectively.}
\end{figure}

We take the vertical distribution of the gas, $n_g(z)$,
from Dickey \& Lockman 1990; the dust distribution, $n_d(z)$, is assumed to be
initially exponential with a scale height $z_d=120$~pc. The initial position and
radius of each (spherical) grain are extracted via a Monte Carlo procedure from
the
parent distributions, $n_d(z)$ and MRN,
respectively.
We then follow the dynamical evolution of the grain ensamble as driven by
radiation, gravity and drag (viscous + coulomb) forces; the grain charge is also
calculated consistently solving the detailed balance equation in which both
collisional and photoelectric charging rates are included.
The grain charge  $Z$ is given by the solution of  

$$n_p\left\{\left({8 k T_e\over \pi m_p}\right)^{1/2}\left[
{\tilde J}_{\nu_p,\tau_p} (a, Z, T_e)s_p + \delta_p(a, Z, T_e)\right]+
{\tilde J}_{pe}(a, {\cal F}_\nu)\right\}=$$
$$n_e \left\{\left({8 k T_e\over \pi m_e}\right)^{1/2}
\left[ {\tilde J}_{\nu_e,\tau_e} (a, Z, T_e)s_e - \delta_e(a, Z, T_e)
\right]\right\},$$
where the index $p$,$e$ stands for protons and electrons, respectively;
${\tilde J}_{\nu,\tau}$  are the collisional charging rates;
$\delta$  are the secondary emission rates; ${\tilde J}_{pe}$ is the
photoelectric charging rate; $s$ is the sticking probability.
${\tilde J}_{\nu,\tau}$ are taken from Draine \& Sutin (1987);
${\tilde J}_{pe}$ is taken from Draine (1978); $\delta$ and $s$ are
taken from Draine \& Salpeter (1979).
The detailed balance equation above  must be solved together 
with the field emission condition (Draine
\& Sutin 1987) that limits the value of the charge:
\begin{equation}
-1-0.7\left({a\over{\rm nm}}\right)^2< Z < 1 + 21\left({a\over{\rm
nm}}\right)^2.
\end{equation}

\begin{figure}
\vspace{10cm}
\caption{Spatial distribution of dust grains for an association (located at the
origin of the axes) containing $N=500$ OB stars at the evolutionary times
shown in the upper corner of each panel}
\end{figure}

Fig. 2 illustrates the spatial distribution of the dust grains (all sizes
$a=10-250$~nm are shown) for a $N=500$ OB stars association at various
evolutionary times.
An inspection of Fig. 2 shows the presence of a central cavity
completely devoided of dust swept by the very strong radiation
pressure. In addition, the configuration of the
region affected by the radiation field resembles now an inflated bubble
in which a relevant fraction of grains has
been pushed well out of the main disk of the Galaxy, at $\vert z \vert >
3$~kpc; it is also possible to identify the
filamentary structures that form the lateral walls of the dusty chimney.
A vertical cut through the origin (i.e. above
the center of the association) shows that the $z$-distribution of the dust
becomes much flatter with time than the original exponential one.
Thus, the dust is rather homogeneously distributed by the
radiation pressure in the portions of the halo above the associations; moreover,
this implies that the dust-to-gas ratio has to be larger in the upper halo than
in the disk. Finally, since the drag force becomes vanishingly small because the
gas density is rapidly decreasing with height, some grains can continue their
journey for several kpc before they feel the gravitational pull of the Galaxy.
The velocities achieved by the grains are rather high, and
they can exceed (for a $N=500$ association) $100$~km~s$^{-1}$. Smaller grains
tend to move faster since - neglecting drag forces - the ratio of the
radiation-to-gravitational force is $\propto a^{-1}$.

\section{A Possible Scenario}

If dust is present in galactic halos it could be responsible for several
important detectable effects. In addition to the standard direct observations
discussed in the Introduction and involving either the obscuration of
distant object by the grains or their FIR emission, a number of indirect but
nonetheless intriguing consequences of this component can be individuated and
used to put constraints on any theory modelling of dusty halos.

\begin{figure}
\vspace{11cm}
\caption{Schematical representation of the various processes occurring above
a HII chimney discussed in the text}
\end{figure}

The simple sketch presented in Fig. 3 helps building a possible scenario
which might be quite common among star-forming spiral galaxies at any epoch  
and includes both the effects discussed above and additional ones.
The gaseous disk of the Milky Way is now understood to consist of at least  
two components: a neutral one, which extends roughly up to $z\sim 0.5$~kpc 
(including the so-called Lockman layer), and an ionized component (Reynolds
layer) whose vertical extent is less clear but of the order of 1 kpc. 
An ordered  magnetic field component with strength $\vert {\bf B}\vert
\sim 3 \mu$G in the
plane of the disk is known to be present (Heiles 1995). 
OB associations will produce (before SN explosions), as we have seen above,
a strong ionizing flux and a HII chimney, pushing the grains into the halo   
by radiation pressure. Due to the presence of the magnetic field, 
and due to the higher density of the gas and
hence enhanced drag, grains located initally at low-$z$ will be prevented 
from being evacuated, {\it if they have a non-zero charge}. This last
point is very important, since in spite of the fact that we have not included
the Lorentz force in our simulations, nevertheless our accurate determination
of the grain charge shows that considerable time intervals are spent by grains
in a neutral charging state: this is due to the balance between photoelectric
and collisional charging.  Thus, at least intermittently, grains will be free 
to move even in presence of a magnetic field above HII chimneys.       
During the periods in which they are charged, grains are trapped and spiral
around field lines. However, grains located at higher altitudes ($z\sim
0.3-0.4$~kpc) will experience a much reduced drag force and therefore the 
pressure of the dust "fluid" will be felt by the magnetic field. This pressure
adds up with the standard thermal and cosmic rays one and will enhance the
Parker instability, thus originating a vertical component of ${\bf B}$. 
The criterion for instability when a dust pressure term $\delta=p_r/p_{th}$
is added to the magnetic field, $\alpha=p_B/p_{th}$, and cosmic ray,
$\beta=p_{cr}/p_{th}$ pressure (where $p_{th}$ is the thermal pressure of the
gas) becomes in the long wavelength limit: 
\begin{equation}
\gamma < {(1+\alpha+\beta+\delta)^2\over 1+\beta+\delta + (3/2)\alpha},
\end{equation}
where $\gamma=5/3$ is the ratio of the specific heats. Since $\delta \gg 1$
above HII chimneys, radiation pressure can profoundly modify the topology
of the magnetic field. Grains spiralling along the vertical field lines
can now flow into the halo and eventually be sputtered (nonthermal sputtering
is likely due to the high grain velocities), releasing the refractory
elements detected by absorption lines experiments through the halo and
know to have a scaleheight of the order of a few kpc, depending on the
species (Edgar \& Savage 1989; Lipman \& Pettini 1995; see Ferrara et al. 1991
for a detailed calculation). This implies that the 
size distribution of grains can deviate substantially from a MRN distribution
in the halo, with an excess of small grains. Finally,  collisions with the gas
atoms will put grains into rotation and, being charged, they will 
emit radiation with a spectrum sharply peaked at a cutoff frequency 
typically in the range 10-100~GHz (Ferrara \& Dettmar 1994). 
The estimate of the global dust mass loss due to this effect from
galactic disks depends on the detailed modelling of the star formation
rate in the galaxy and other galactic properties and will be given in 
Ferrara \& Shull (1997).

\section{Cosmological Implications of Dusty Halos}

There are several consequences of the presence of dust in the halos
of galaxies and perhaps in the intergalactic medium. The most obvious one
concerns the determination
of the deceleration parameter, $q_0$; indeed, this parameter is extremely
sensitive
to the assumptions concerning the extinction. In an important paper, Margolis
\& Schramm (1977) have estimated that the correction due to EGD to $q_0$ is
$$(\Delta q_0) = 2.8\times 10^3 \left({A_v\over 1~{\rm mag~Mpc}^{-1}}\right)
h^{-1},\eqno(1)$$
where $A_v$ is the visual extinction and $h=H_0/100~{\rm km~s}^{-1} {\rm
Mpc}^{-1}$.
Thus, as little as $2\times 10^{-4}~{\rm mag~Mpc}^{-1}$ absorption would cause
a measurement error of $(\Delta q_0)\sim 1$.

Extragalactic dust (i.e. dust outside the main body of galaxies) might also 
affect QSO number counts (Heisler \& Ostriker 1988). The ratio of
the total cross section of dusty absorbers to the total sky area in a $\Omega=1$
universe at redshift $z$ is 
\begin{equation}
{\cal F}(z) = {2 \pi R(0)^2 n(0)c\over 3 H_0}(1+z)^{3/2} 
\end{equation}
where $n(0)$ and $R(0)$ are the local number density and size of the
absorbers. Substituting typical numbers for ($n(0)\sim r_0^{-3}\sim
0.008 h^3$~Mpc$^{-3}$, where $r_0\sim 5 h^{-1}$~Mpc is correlation 
scale for galaxies and $R(0)\sim 100$~kpc) halos 
associated with galaxies, we conclude that ${\cal F}(z)=1$ at redshift
$(1+z)=(2/h^2)^{2/3}=1.58 h^{-1/3}$. Thus high redshift objects are 
likely to suffer substantial extinction by intervening systems.
As a final remark, the release of heavy elements following grain sputtering
in the halo might be relevant to the interpretation of QSO absorption lines in   
metal line systems.

\bigskip
I deeply acknowledge the constant inspiration provided by the friend 
Prof. Lyman Spitzer, Jr. ``from whom I have learned a lot''. 
I also thank Mike Shull, my collaborator in this project.

%
%

\end{document}